\documentclass[twocolumn]{aastex631}

\usepackage{amsmath}
\usepackage{units}
\usepackage{graphicx}
\usepackage{subfigure}
\usepackage{lmodern}
\usepackage{comment}
\shorttitle{Properties of \ion{Mg}{2} lines in Flares}
\shortauthors{Roy and Tripathi}

\begin{document}

\title{Evolution of the Ratio of \ion{Mg}{2} Intensities During Solar Flares}
\author[0000-0003-2215-7810]{Soumya Roy}
\affil{Inter-University Centre for Astronomy and Astrophysics, Post Bag 4, Ganeshkhind, Pune 411007}
\affil{Harvard-Smithsonian Center for Astrophysics, 60 Garden Street, Cambridge, MA 02138, USA}
\author[0000-0003-1689-6254]{Durgesh Tripathi}
\affil{Inter-University Centre for Astronomy and Astrophysics, Post Bag 4, Ganeshkhind, Pune 411007}

\correspondingauthor{Soumya Roy}
\email{soumyaroy@iucaa.in}

\begin{abstract}

The \ion{Mg}{2}~k \& h line intensity ratios can be used to probe the characteristics of the plasma in the solar atmosphere. In this study, using the observations recorded by the Interface Region Imaging Spectrometer (IRIS), we study the variation of the \ion{Mg}{2}~k \& h intensity ratio for three flares belonging to X-class, M-class, and C-class, throughout their evolution. We also study the k-to-h intensity ratio as a function of magnetic flux density obtained from the line-of-sight magnetograms recorded by the Helioseismic and Magnetic Imager (HMI) on board the Solar Dynamics Observatory (SDO). Our results reveal that while the intensity ratios are independent of magnetic flux density, they show significant changes during the evolution of the C-class and M-class flares. The intensity ratios start to increase at the start of the flare and peak during the impulsive phase before the flare peak and decrease rapidly thereafter. The values of the ratios fall even below the pre-flare level during the peak and decline phases of the flare. These results are important in the light of heating and cooling of localized plasma and provide further constraint on the understanding of flare physics.

\end{abstract}

\keywords{Sun: chromosphere -- Sun: flares -- Sun: UV radiation}

\section{Introduction} \label{sec:intro}

Solar flares are the most energetic events on the Sun, where an enormous amount of magnetic free energy is released due to the reconfiguration of the coronal magnetic field. The released energy can cause particle acceleration, heating and flows in the solar atmosphere and a transient enhancement in solar radiative output. It is observed that most of the energy radiated in flares originates from the dense chromosphere \citep{fletcher10,milligan14}. Hence, studying the chromospheric lines during flares provides us with diagnostics, which may be important for understanding the physics of solar flares and their effect on the local plasma environment.

The chromosphere emits in various UltraViolet(UV) and optical lines. While many optical lines, e.g., H$\alpha$, \ion{Ca}{2}, are routinely observed from ground-based telescopes, observations in the \ion{Mg}{2} resonance lines have been rare in the past. Since the launch of the Interface Region Imaging Spectrograph \citep[IRIS;][]{IRIS} we have been in the position to monitor these lines regularly with excellent spatial and spectral resolution.

The \ion{Mg}{2}~k and h lines are transitions to the ground state from a finely split pair of upper levels ($3p~^{2}P_{\nicefrac{3}{2}}${--}$3s~^{2}S_{\nicefrac{1}{2}}$ and $3p ^{2}P_{\nicefrac{1}{2}}${--}$3s~^{2}S_{\nicefrac{1}{2}}$). These transitions create the optically thick lines in the wavelengths 2796.34~{\AA} (\ion{Mg}{2}~k) and 2803.52~{\AA} (\ion{Mg}{2}~h). It is suggested that the intensity ratios of these lines can be used to probe the optical depth of the local environment \citep{kerr15}. 

The integrated intensity of a line transitioning from an upper-level j to a lower-level i, is dependent on the collision strength for that transition $\Omega_{ij}$, which is given by~\citep{henri62,mariska92},

\begin{equation*}
    \Omega_{ij}~=~\frac{8\pi}{\sqrt{3}}~\frac{I_{H}}{\Delta \epsilon_{ij}}~g~\omega_{i}~f_{ij}
\end{equation*}

\noindent Where $I_{H}$ is the ionization energy of Hydrogen, $\Delta \epsilon_{ij}$ is the threshold energy for the transition, g is the Gaunt factor, $\omega_{i}$ is the statistical weight of the level and $f_{ij}$ is the oscillator strength. In optically thin conditions, the intensity ratio of the k to h line is the ratio of the collision strengths, as the escape probability of photon is unity. The \ion{Mg}{2} k and h lines share the same ionization state and originate from a transition to a shared lowered level. As the statistical weight ($\omega_{i}$) is same in both cases, the line intensity ratio is simply the ratio of the oscillator strengths($f_{ij}$). This implies the ratio is 2:1 in optically thin conditions, and lower when the medium is optically thick \citep{kerr15,levens19}.

In addition, the \ion{Mg}{2} k and h lines can be used to estimate the velocity in the middle and upper chromosphere, the chromospheric velocity gradients, the temperature in the middle chromosphere \citep{leenarts13a,leenarts13b,pereira13}. The \ion{Mg}{2} triplets in emission can be used to identify heating in the lower chromosphere \citep{pereira15}. There have also been multiple studies that have shown a spatial variation of the \ion{Mg}{2} line profiles \citep{dalda23,panos18}. \cite{polito23} showed that the leading edge of the flare ribbon is associated with enhanced broadening and strong central reversal. They interpreted the difference in the profile as a difference in the heating mechanism at the leading edge and bright part of the flare ribbons. \cite{panos21,panos21_2} showed similar differences between the line profiles and energy input.

Using the observations recorded by the OSO-8 LPSP instrument, \citep{lemaire84} studied the evolution of the intensity ratio of \ion{Mg}{2}~h~\&~k, \ion{Ca}{2}~h~\&~k and Ly~$\alpha$~\&~$\beta$ lines. \cite{lemaire84} showed that the intensity ratio of the \ion{Ca}{2}~k/h lines increased from 1 to 1.2 during the ascending phase of a flare and returned to 1 during the later phases. The authors interpreted the correlated temporal behavior across various elements as an indication of downward energy propagation. We note that this may suggest a slight decrease in the opacity due to localized heating at the formation height of the \ion{Ca}{2} line during the rise phase of the flare.


Here, we study the evolution of the intensities ratios of the \ion{Mg}{2}~h~\&~k lines during the course of the evolution of three flares, \textit{viz.}, C-class, M-Class and X-class. We focus on the the dependence of line ratios on the underlying magnetic field strength, which to our knowledge has not been explored so far. The rest of the paper is structured as follows. \S\ref{sec:obs} discusses the observations used in this paper. We discuss how we reduce and analyze the data and the results in \S\ref{sec:dar}.

\section{Observations} \label{sec:obs}
\begin{table*}[ht!]
     \centering
 \begin{tabular}{|c|c|c|c|c|c|}
  \hline     
       Event & Flare  & Flare  & Raster & Raster Step & Raster \\
        Date & Peak (UT) & Location (arcsec) & Details & (arcsec) & Cadence (s)\\
       \hline
        Nov 4, 2015 (M-class) & 13:52 & [37",61"] & Coarse 16 step & 2" & 50\\
        Oct 22, 2014 (X-class) & 14:28 & [-292",-302"] & Coarse 8 step & 2" & 131\\
        Feb 3, 2015 (C-class) & 22:55 &  [198",213"] & Dense 16 step & 0.35" & 33\\
        \toprule
    \end{tabular}
    \caption{The list of flares studied in this paper.}
    \label{tab:my_label}
\end{table*}

For this study, we have selected three flares belonging to M, X and C classes, listed in Table~\ref{tab:my_label} by IRIS. IRIS is a NASA small explorer-class solar observation satellite. It obtains UV spectra with high spatial (0.33{--}0.4\arcsec per pixel), temporal (1s), and spectral resolution ($\sim$26 and $\sim$53~m{\AA}). In the spectral channels, the strongest lines that IRIS regularly observes are \ion{C}{2}, \ion{Mg}{2} and \ion{Si}{4}. In the imaging channel, it typically observes in \ion{Mg}{2} and \ion{Si}{4}. In this work, we have used the observations recorded in \ion{Mg}{2}~h~\&~k lines. 

As alluded earlier, the aim of this study is to derive the evolution of intensity ratios as a function of magnetic flux density during the course of flares of varying class. For this purpose, we have used line-of-sight (LOS) magnetic field measurements from the Helioseismic and Magnetic Imager \citep[HMI;][]{hmi} on board the Solar Dynamics Observatory \citep[SDO;][]{sdo}. We have used observations taken at 1600~{\AA} by the Atmospheric Imaging Assembly \citep[AIA;][]{aia}, also onboard SDO, to co-align the IRIS observations with those from AIA and then HMI.

\section{Data analysis and results} \label{sec:dar}
\subsection{M3.7 Flare Observed on Nov 4, 2015}
\begin{figure*}[ht!]
    \centering
\hspace*{-.6in}
\includegraphics[trim={7cm 5cm 7.7cm 6cm},clip,width=1.1\textwidth]{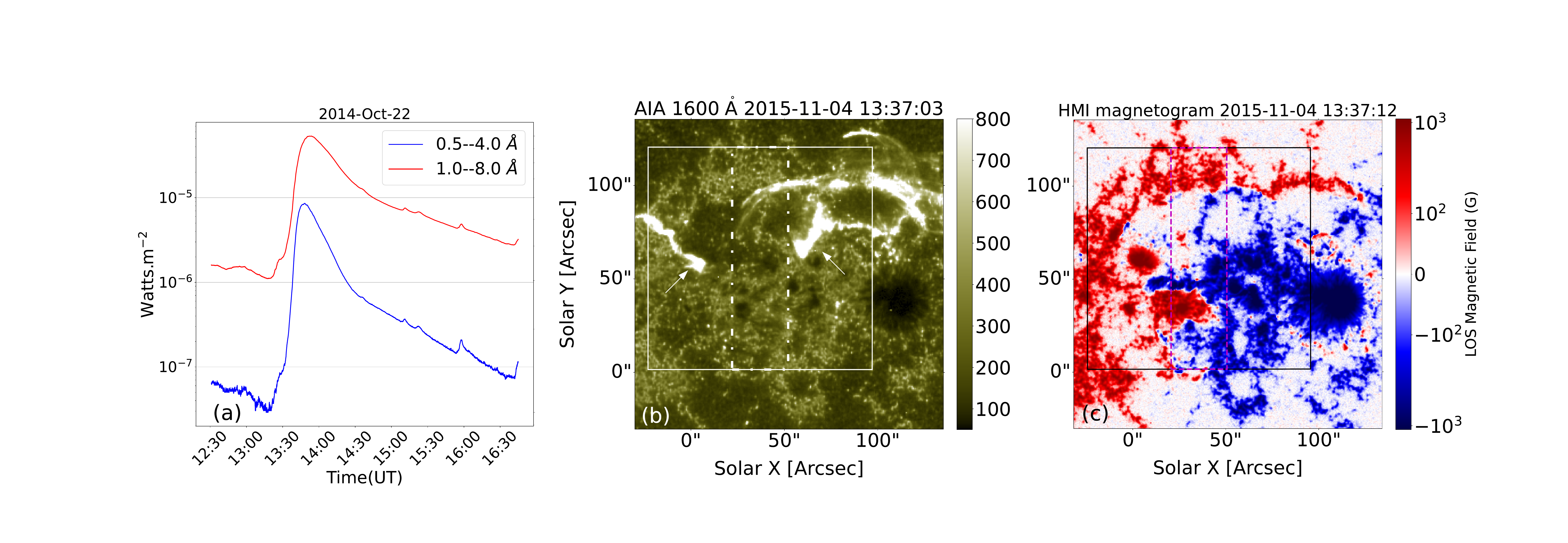}
\caption{The M3.7 flare observed on November 4th, 2015. Panel a: GOES flux plot in 0.5{--}4~{\AA} (blue) and 1.0{--}8.0~{\AA} (red). Panel b: AIA 1600~{\AA} image of the flaring region. Arrows locate the primary ribbons. Panel c: LOS magnetic flux density map obtained from HMI near the peak of the flare. The over-plotted white (black) boxes in panel b (c) represents the IRIS SJI FOV. The over-plotted white dot-dashed (magenta dashed) box in panel b (c) show the IRIS raster FOV.}\label{flare1}
\end{figure*}

The NOAA~AR~12443 produced a multi-ribbon GOES class M3.7 flare on November 4, 2015 that began $\sim$~13:31~UT, and peaked at $\sim$ 13:52~UT as seen from the GOES SXR 1{--}8~{\AA}~flux(Fig.~\ref{flare1}(a)). The event was located at the heliographic position of $\sim$ [37\arcsec,61\arcsec] and very well observed by IRIS, AIA and HMI. Fig.~\ref{flare1}.a displays the GOES flux plot of the flare in 0.5{--}4~{\AA} (blue) and 1.0{--}8.0~{\AA} (red). Fig.~\ref{flare1}.b is an AIA 1600~{\AA} image of the flaring region consisting of two ribbons, indicated by arrows. In panel c, we plot the LOS magnetic flux density map obtained from HMI recorded nearly simultaneously with AIA image shown in panel b. The white (black) box over plotted in Fig.~\ref{flare1}.b (c) show the IRIS SJI FOV. The over-plotted white dot-dashed (magenta dashed) box in Fig.~\ref{flare1}.b (c) represents the IRIS raster FOV. The FOV of the SJI ($\sim~[120\arcsec~\times~119\arcsec]$) covers the central part of the flaring region. The spectral sampling is $\sim$~0.05~{\AA}/pixel. The dynamics of the ribbons for this flare is studied by \cite{li17}, while \cite{karlick18} studied the associated radio bursts.

\begin{figure*}[ht!]
    \begin{center}
        \begin{interactive}{animation}{nov_flare_304_aa_evolv.mp4}
        \includegraphics[width=0.9\textwidth]{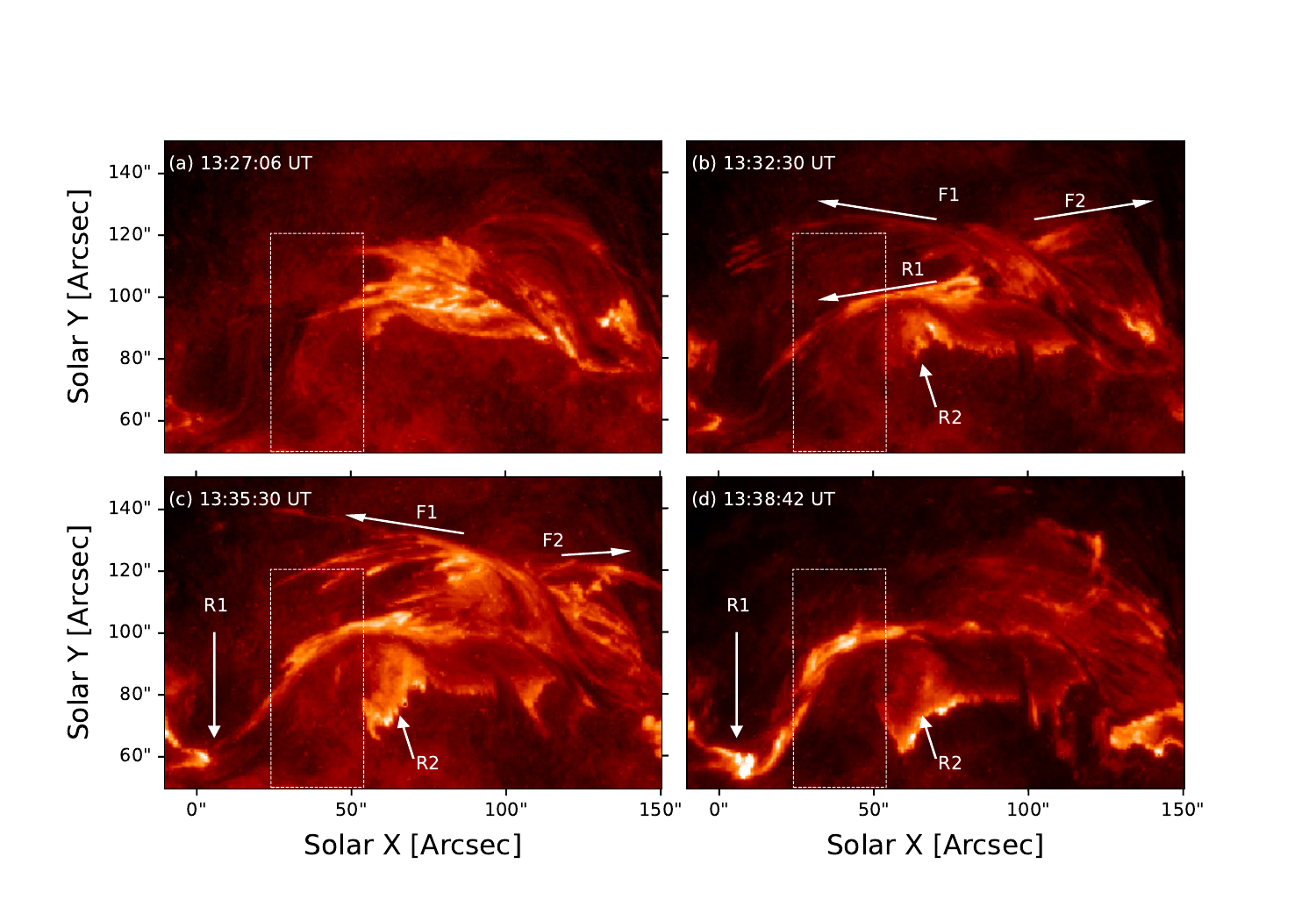}
        \end{interactive}
    \end{center}
    \caption{Sequence of AIA 304~{\AA}~images for the Nov 4th, 2015 flare. The white dotted box shows the portion of the IRIS raster FOV which scanned the ribbons. In panels b (c) F1 and F2 are the filament material, which move away from each other as the flare progresses. R1 and R2 in b (c,d) show the primary ribbons which move through the IRIS raster FOV. An animation of this image sequence is available in the online version. The animated images do not contain the annotations for the filament and ribbon. The IRIS raster FOV is annotated with a white dotted box. The animation runs from 13:26:18 {--} 13:39:54 UT.}
    \label{flare_m_ev}
\end{figure*}

Fig.~\ref{flare_m_ev} shows the evolution of the flare in AIA 304~{\AA}. The flare is associated with a pre-existing filament that erupts and breaks into two structures, marked as F1 \& F2 in Fig.~\ref{flare_m_ev}.b~\&~c. These two filament structures moved away from each other in opposite directions. The flare produces two primary flare ribbons, marked as R1 \& R2 in Fig.~\ref{flare_m_ev}.b,~c~\&~d. From $\sim$ 13:32 UT, R1 moves southeast as it sweeps across the IRIS raster FOV, marked by the white dotted box in Fig.~\ref{flare_m_ev}.a,~b,~c~\&~d. The IRIS raster scans the motion of the northern ribbon R1 and the eastern edge of R2. An animated version of the Fig.~\ref{flare_m_ev} is available in the online journal for further details.

\begin{figure*}[ht!]
    \begin{center}
        \includegraphics[trim={4.5cm 5.5cm 6cm 4cm},clip,width=0.9\textwidth]{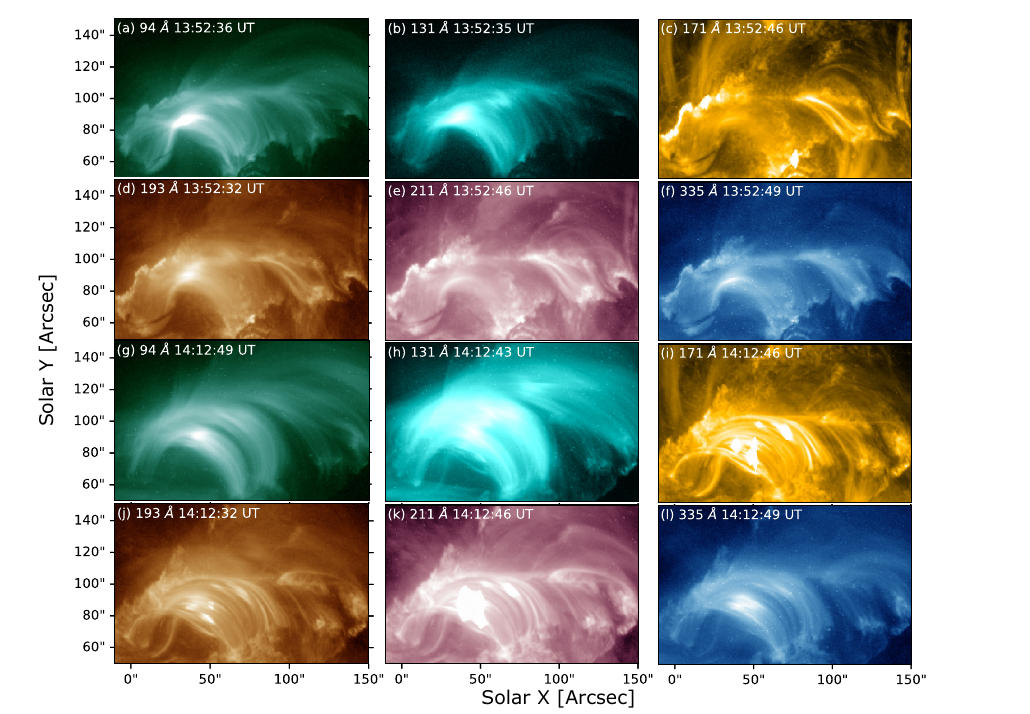}
    \end{center}
    \caption{The Nov 4th, 2015 flare in the six Coronal channels of the SDO/AIA at the soft X-ray peak (panels (a)-(f)) and during the decline phase (panels (g)-(l)) of the flare.}
    \label{flare_m_aia}
\end{figure*}

Fig.~\ref{flare_m_aia} displays the same region as in Fig.~\ref{flare_m_ev} in the six coronal channels (namely 94, 131, 171, 193, 211, 335~{\AA}) of SDO/AIA, recorded at the peak (top two rows) and during the decline phase (bottom two rows) of the flare. Post-eruption arcades \citep[][]{TriBC_2004} are clearly observed in all the channels with slightly different morphology. These arcades are filled with the evaporated thermal plasma, and also show loop top brightening, likely due to the colliding evaporation flows \citep[see, e.g.,][]{sharma16, patsourakos04}.

We first co-align the HMI observations with the IRIS observations, using the 1600~{\AA} recorded by AIA. Since the observation is of an active region that is undergoing flaring, the magnetic field may also be changing rapidly \citep{wang02,dandan16,spirock02}. Therefore, we use a series of AIA and HMI full disk co-aligned maps to obtain rastered LOS map of magnetic flux density that corresponds precisely to the location and time of IRIS rasters. This procedure is detailed below.

Initially, we align the AIA 1600~{\AA}~observation with the HMI observation, with the closest observation time to the IRIS SJI and raster observations. Using `aia\_prep' available in the \textit{sswidl} distribution, we process the level 1 images to perform image registration and align AIA and HMI observations. Thereafter, we compute the shift between AIA 1600~{\AA} and the SJI 1400~{\AA}. Using the calculated offsets, the magnetograms are coaligned with the IRIS SJI 1400~{\AA} observations, since the AIA 1600~{\AA} observation was already aligned with the HMI observations.We find the typical offset between HMI and IRIS is $\sim$ 1.5\arcsec. We use these co-aligned HMI maps to obtain the rastered magnetograms, which could then be directly compared with IRIS raster observations. 

For characterizing the properties of the \ion{Mg}{2} lines, we fitted a double Gaussian profile to both k and h lines with a linear background symmetric about the line core. If excess emission is detected compared to the fitted background and line profile for wavelengths lower than 2792~{\AA} and in-between 2798~{\AA} to 2800~{\AA}, we assume that the \ion{Mg}{2} triplets have appeared in emission. A single Gaussian profile is fitted to the excess emission. We note that this is prone to missing the triplets unless the emission is reasonably strong. However, we note that since our main objective was to characterize the \ion{Mg}{2} k and h line profiles, this procedure serves our purpose. We also excluded any pixels that showed saturation in either of the lines. The uncertainty of the observed intensities are measured in DN by the method detailed in \S 2.1 in \cite{kerr15} and used subsequently in the fitting procedure. The uncertainties of the fitted Gaussian profiles are propagated while integrating the profile, to obtain the uncertainty of the line intensities.

\begin{figure}[ht!]
\centering  
\includegraphics[trim={7cm 5cm 5cm 3cm},width=0.5\textwidth]{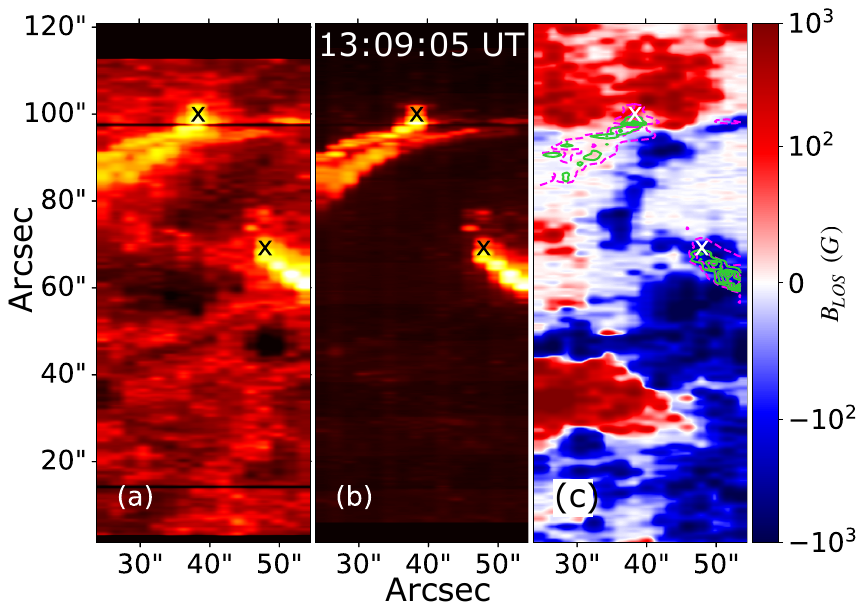}
\caption{Obtained images in \ion{Mg}{2} h (panel a) and k (panel b) and corresponding co-aligned and artificially rastered HMI LOS magnetic field map (panel c) for the M3.7 flare observed on November 4th, 2015. The magenta and lime green contours on panel c show the contours of \ion{Mg}{2}~h (panel a) and k (panel b) intensity.} \label{fig:aligned_raster}
\end{figure}

\begin{figure*}[ht!]
    \centering
    \includegraphics[trim={0cm 3cm 0.5cm 3cm},clip,width=\textwidth]{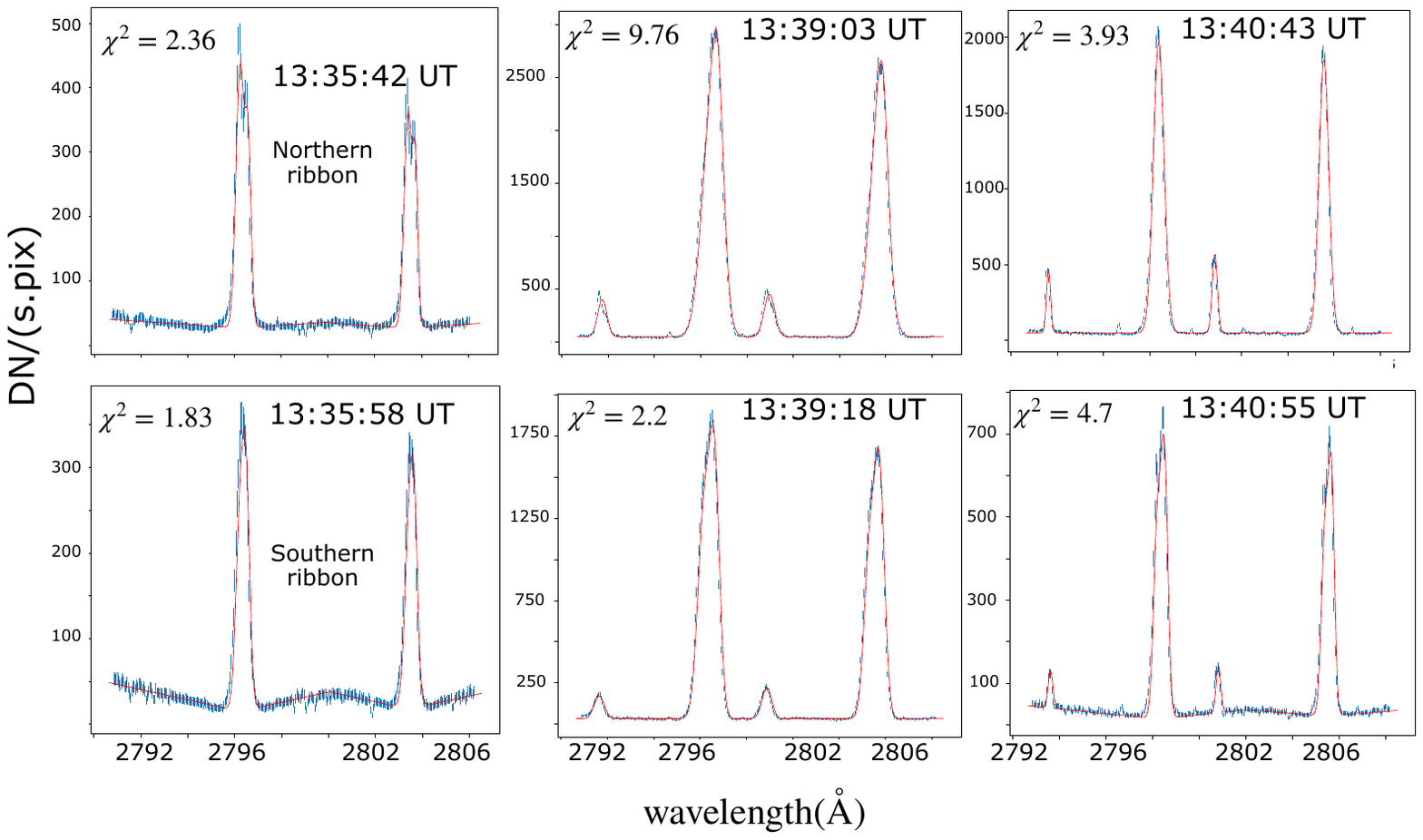}
    \caption{Fit of the \ion{Mg}{2} window for the pixel marked in northern ribbon (southern ribbon) in Fig.~\ref{fig:aligned_raster} in top panel (bottom panel) from various time during the evolution of the flare. }
    \label{fig:pix_fit_ribbon}
\end{figure*}

\begin{figure}[ht!]
\centering  
\includegraphics[trim={1.5cm 4cm 0cm 4cm},clip,width=0.5\textwidth]{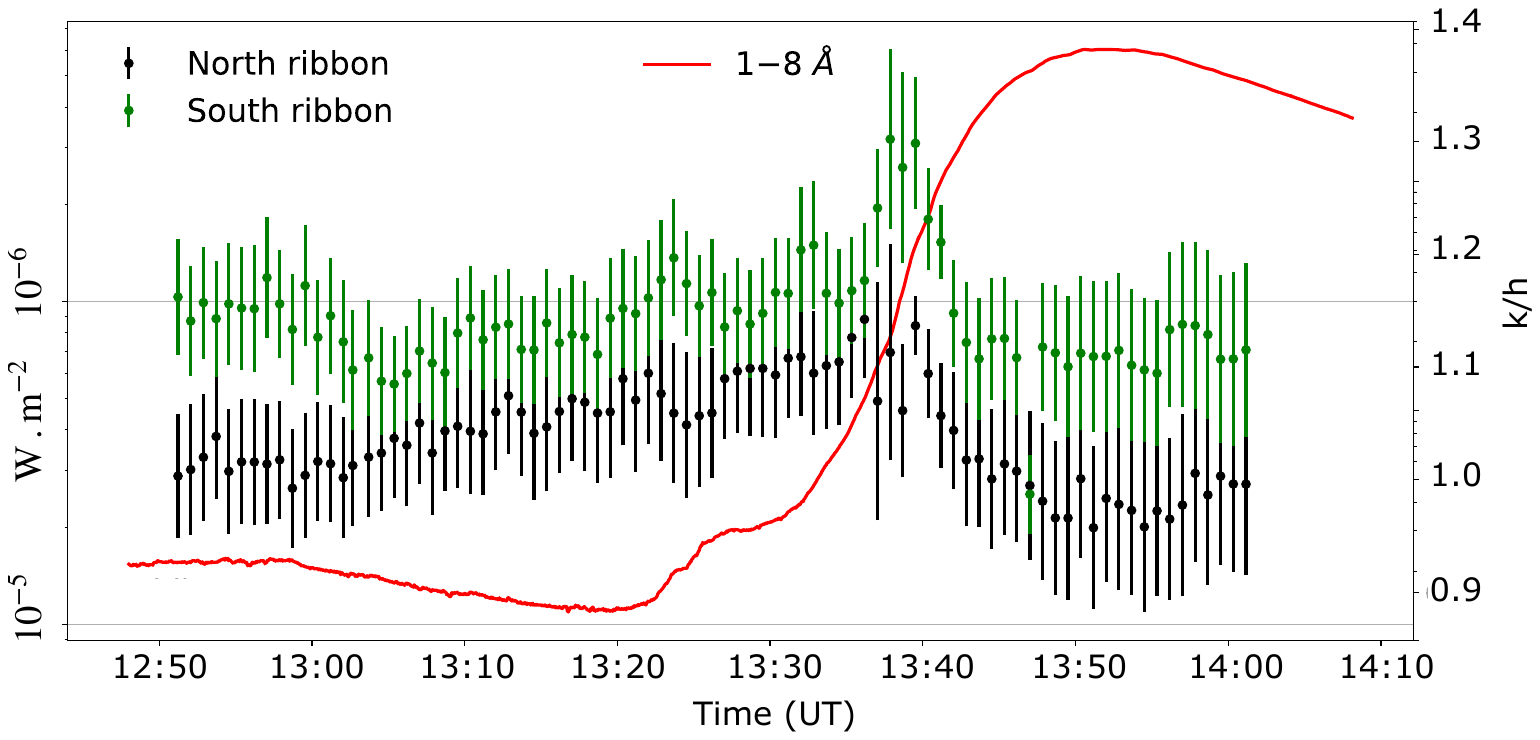}
\caption{The 1{--}8~{\AA} GOES light curve of the flare over-plotted with the time variation of \ion{Mg}{2} k/h line intensity ratio for the northern (black) and southern (green) asterisks marked in panels a, b~\&c.}
\label{fig:aligned_iris_ratio}
\end{figure}

In Figure~\ref{fig:aligned_raster}, we plot the intensity maps obtained in \ion{Mg}{2}~h \& k (panels a \& b) and the rastered LOS magnetic field map in panel (c). The dashed magenta and solid green contours on the magnetic field maps show the intensity from panels a and b, respectively. To study the \ion{Mg}{2} k to h intensity line ratios as a function of time, we chose two pixels -- one in the northern ribbon and another in the southern ribbon, as denoted by crosses. In Fig.~\ref{fig:pix_fit_ribbon}, we display the spectra over-plotted with fits for the two pixels taken in the northern ribbon (top panel) and southern ribbon (bottom panel) obtained at different times.

We plot the \ion{Mg}{2} k to h line intensity ratio obtained from the two locations in the northern (black) and southern ribbons (green) in Fig.~\ref{fig:aligned_iris_ratio}. We also over-plot GOES light curve of the flare obtained in 1.0{--}8.0~{\AA} (red solid line). Our observations reveal that the intensity ratios show a slightly increasing trend, albeit within the errors, during the early phase of the of flare. The ratio displays a distinct peak almost at the mid-way in the impulsive phase, lasting for a very short time $\sim$ 150s, before it starts to decrease. We emphasize here that the pixel-by-pixel fitting of the \ion{Mg}{2}~h~\&~k line profiles provided us with adequate signature of the time evolution of the k/h line intensity ratio during flare. 

\subsubsection{Investigating the dependence on the local magnetic field in the \ion{Mg}{2} lines}

To investigate the time evolution of the intensity ratio and its relationship with the photospheric magnetic field, if any, we binned the spectra in various magnetic field bins from various flaring pixels. We selected the flaring pixels with an intensity threshold with respect to the peak intensity observed in IRIS rasters. We perform a double Gaussian fit with a constant background in a smaller wavelength window for both \ion{Mg}{2} h and k lines separately. This is done essentially to avoid the effects of the background and possible contribution of any other spectral feature.

\begin{figure*}[ht!]
    \begin{center}
    \includegraphics[trim={0cm 3cm 0cm 3cm},clip,width=\textwidth]{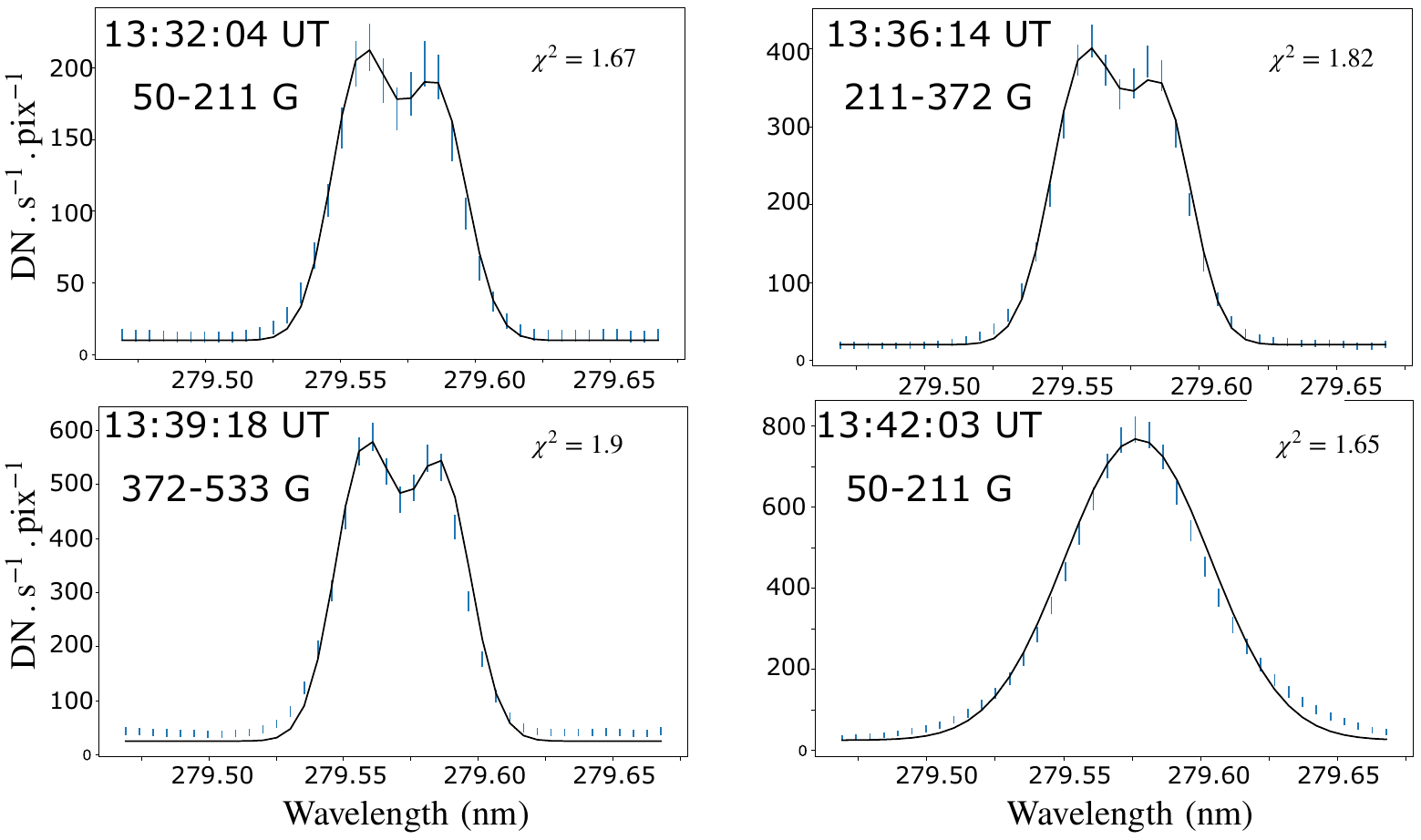}
    \end{center}
    \caption{Fitted line profiles for the binned spectra for various magnetic field strength from various times.}
    \label{fig:fit_pix_fov}
\end{figure*}

We display the \ion{Mg}{2}~k line profiles taken at different times and averaged over different magnetic field bins in Fig.~\ref{fig:fit_pix_fov}. In each plot, we note the time of observation and the magnetic field strength over which the spectra is averaged and fitted double-Gaussian. We compute the line intensities by integrating the fitted Gaussian.

\begin{figure}[ht!]
    \centering
    \includegraphics[trim={8cm 1cm 2cm 0.4cm},clip,width=0.5\textwidth]{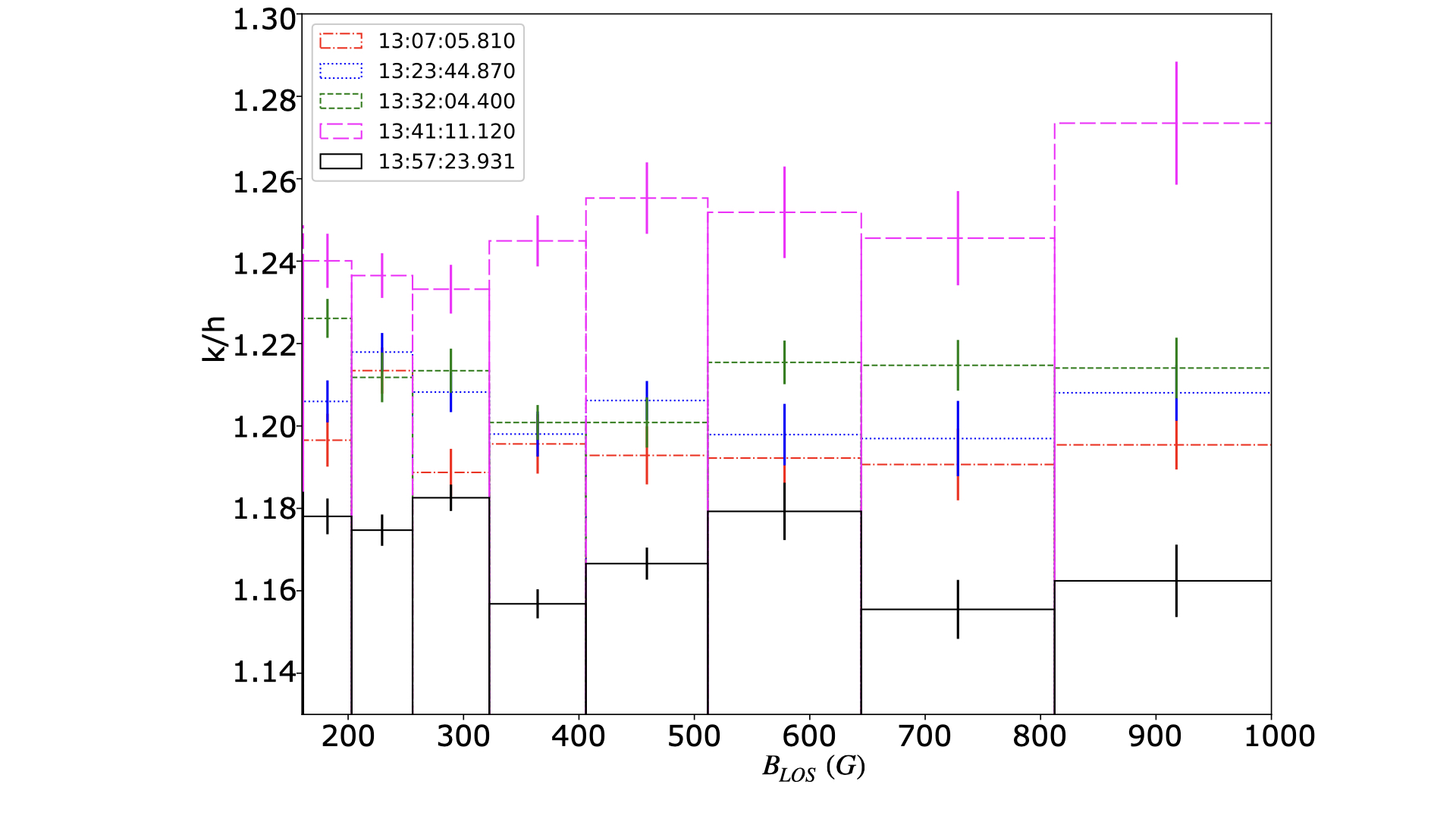}
    \caption{\ion{Mg}{2} k to h line intensity ratio for various magnetic flux density bins at various time steps during the evolution of the flare.}
    \label{fig:optical_depth_m}
\end{figure}

In Fig.~\ref{fig:optical_depth_m}, we plot the k to h intensity ratios obtained in various bins of the magnetic flux density at different times corresponding to different phases of the flare. The plot corresponding to 13:07:05~UT (red curve) is during the pre-flare phase, where the ratio is $\sim$1.2. During the impulsive phase, i.e., the curves corresponding to 13:23:44~UT (blue dotted) and 13:32:04~UT (green dashed), the ratio increases. The ratios measured at 13:41:11~UT (magenta dashed), which is closest to the UV peak of the flare, is the largest across all magnetic field bins. Finally, during the decline phase, the ratio measured at 13:57:23~UT (black solid), the ratio falls below the values of the pre-flare phase. We also note that there are no significant differences among the k to h ratios obtained in different bins of magnetic flux density at all times (including pre-flare, impulsive, peak and decline phases) of the flare.

\begin{figure*}[ht!]
    \centering
    \includegraphics[trim={3cm 3cm 2cm 4cm},clip,width=\textwidth]{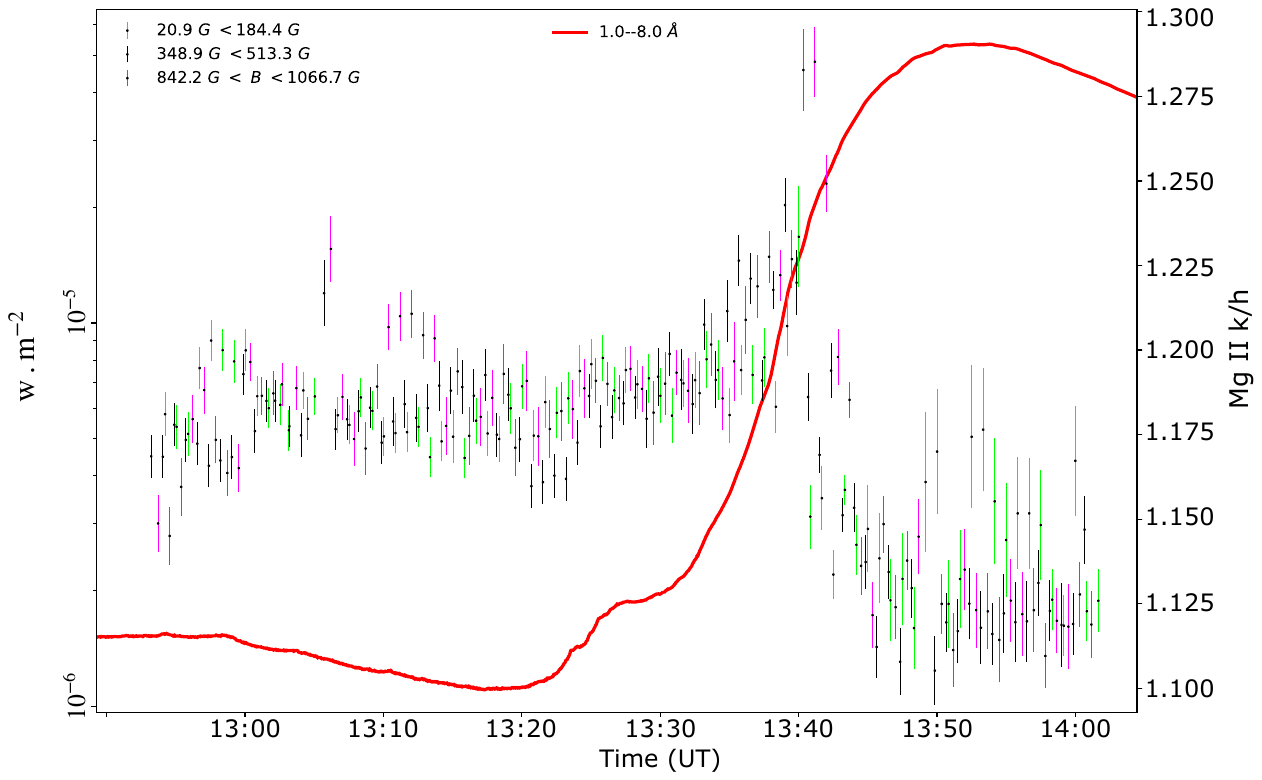}
    \caption{Time evolution of the \ion{Mg}{2} k to h line intensity ratio obtained from averaged spectrum over the corresponding magnetic flux bin as labeled for the Nov 4, 2015 flare. For better visibility the 20.9{--}184.4 G and 348.9{--}513.3 G points are offset by 30s and -30s respectively. Over-plotted red solid line displays the 1{--}8~{\AA} GOES X-ray light curve.}
    \label{fig:optical_dep_ev_m}
\end{figure*}

To explore further, we study the evolution of k to h intensity ratio as a function of time during the course of the flare. In Fig.~\ref{fig:optical_dep_ev_m}, we plot the time evolution of the k to h line intensity ratio averaged within the bins of different magnetic flux densities as a function of time. For better visibility the 20.9{--}184.4 G and 348.9{--}513.3 G points are offset by 30s and -30s respectively. We also plot the GOES 1{--}8~{\AA} light curve with red solid line. These plots conspicuously reveal that the intensity ratios rise sharply during the impulsive phase, from $\sim$ 1.20 to $\sim$ 1.28 right before the soft X-ray flux peaks as seen from GOES, and decreases very rapidly thereafter, to lower than pre-flare values $\sim$ 1.12. The typical value of the uncertainty for the line intensity ratio is $\sim$ 0.02. We further note that the evolution of the \ion{Mg}{2}~ k to h intensity ratio is remarkably similar across various strength of magnetic flux densities. 

We have also carried out this analysis by obtaining the intensity ratios using other methods and note that the obtained results are very similar, see Appendix~\ref{appendix}.

\subsection{X1.6 Flare Observed on Oct 22, 2014}

\begin{figure*}[ht!]
    \centering
\hspace*{-.5in}
\includegraphics[width=1.12\textwidth,trim={7cm 5cm 7.7cm 6cm},clip]{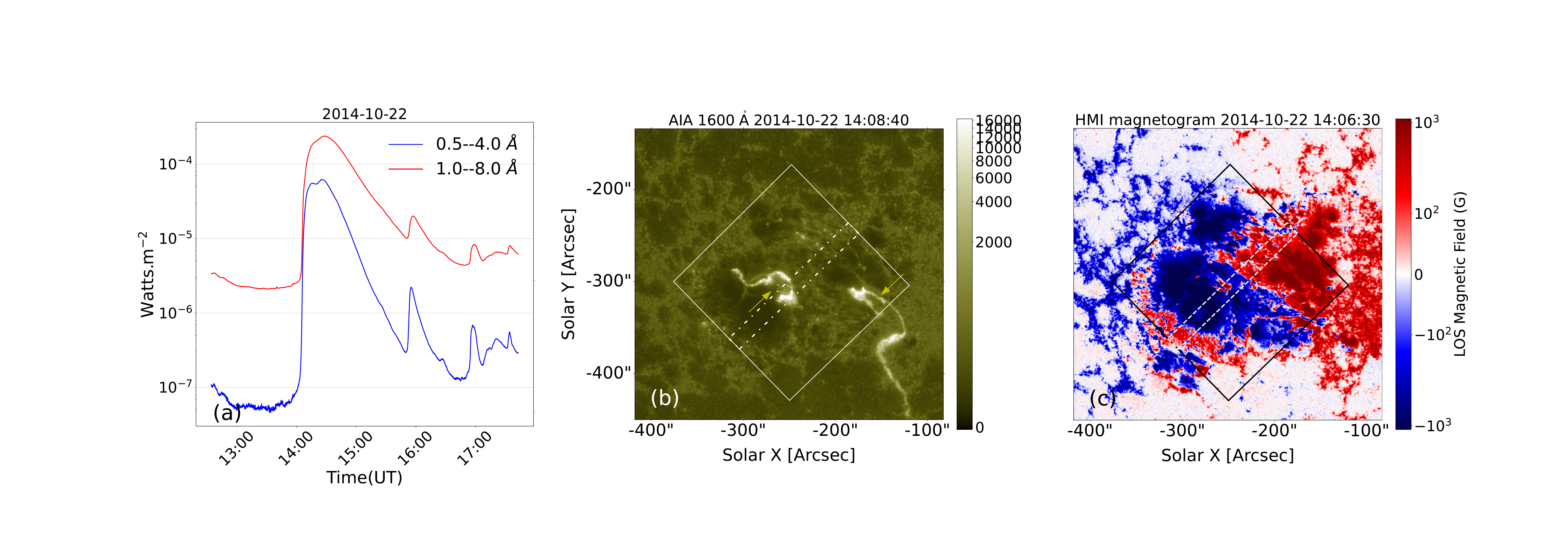}
\caption{The X class flare observed on October 22, 2014. Panel a: GOES flux plot in 0.5{--}4~{\AA} (blue) and 1.0{--}8.0~{\AA} (red). Panel b: AIA 1600~{\AA} image taken at the peak of the flare. Arrows locate the primary and secondary ribbons. Panel c: LOS magnetic flux density map obtained from HMI at the peak of the flare. The over-plotted white (black) box in panel b (c) represents the IRIS SJI FOV, and white dot-dashed (dashed) box in panels b (c) represents the IRIS Raster FOV.}\label{flare2}
\end{figure*}

An X~1.6 flare on October 22, 2014 was observed in AR~12192. The flare started at 14:02~UT, and peaked at 14:28~UT as observed in by GOES. Fig.~\ref{flare2}.a displays the GOES flux plot of the flare in 0.5{--}4~{\AA} (blue) and 1.0{--}8.0~{\AA} (red). Figs.~\ref{flare2}.b \& c display AIA 1600~{\AA} image and LOS magnetic flux density map obtained from HMI, respectively. The over-plotted boxes in panels b and c show the IRIS SJI FOV. IRIS observed this flare with a large 8-step coarse raster covering a FOV of [14\arcsec,174\arcsec]. Note that, the IRIS SJI FOV and also the slit direction was rotated $\sim$ 45$^\circ$ relative to the center of HMI observation.

\begin{figure}[ht!]
    \centering
    \includegraphics[trim={6cm 3cm 6cm 3cm},clip,width=0.5\textwidth]{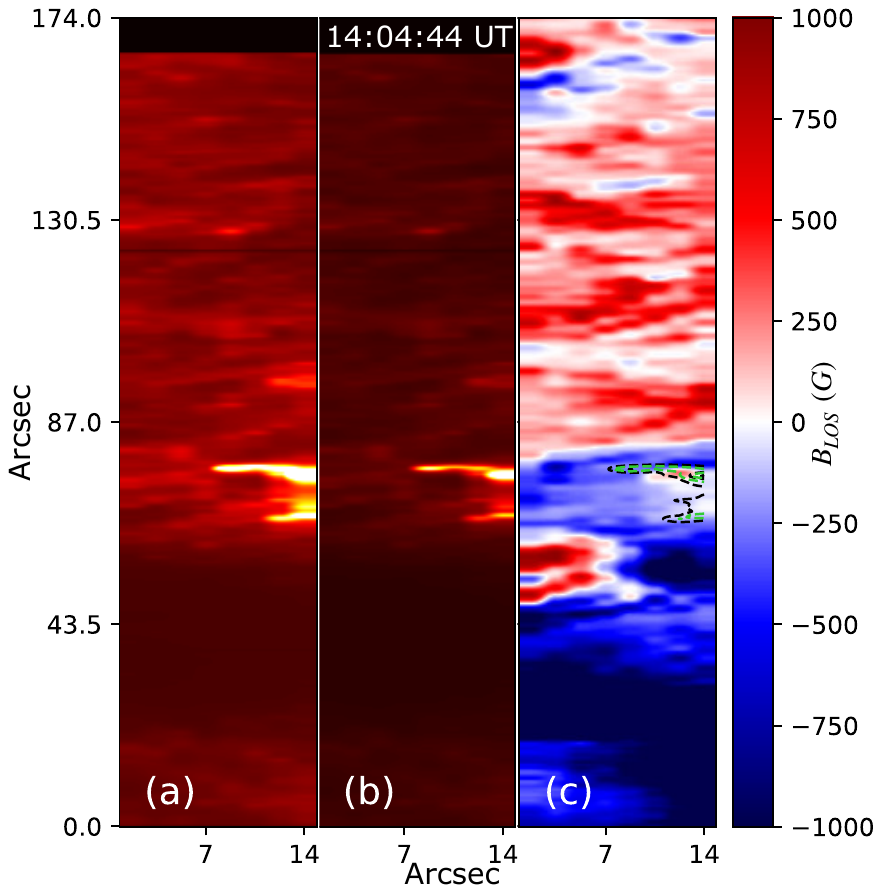}
    \caption{Obtained intensity maps in \ion{Mg}{2}~k (panel a) and h (panel b) lines at the peak of the flare. The rastrered HMI LOS magnetic field density map is shown in panel c. The black and lime green contours on panel c show the contours of \ion{Mg}{2}~k (panel a) and h (panel b) intensity.}
\label{fig:align_raster_flare2}
\end{figure}

The flare exhibits two ribbons in AIA~1600{\AA} as indictaed by the arrows. The eastern ribbon extends around a negative field region and is completely covered by the IRIS SJI FOV. The western ribbon, however, extends around a positive field region and only a part of it is covered by the IRIS SJI FOV, as shown in panel b. In the Fig.~\ref{fig:align_raster_flare2}, we plot the intensity maps obtained in \ion{Mg}{2}~h \& k (panels a \& b), and the rastered LOS magnetic field map (panel c). The black \& green dashed contours in the LOS magnetic field map (panel c) show the intensity contours of \ion{Mg}{2}~h \& k (panels a \& b).

\begin{figure*}[ht!]
    \centering
    \includegraphics[trim={3cm 2cm 2cm 3cm},clip,width=0.8\textwidth]{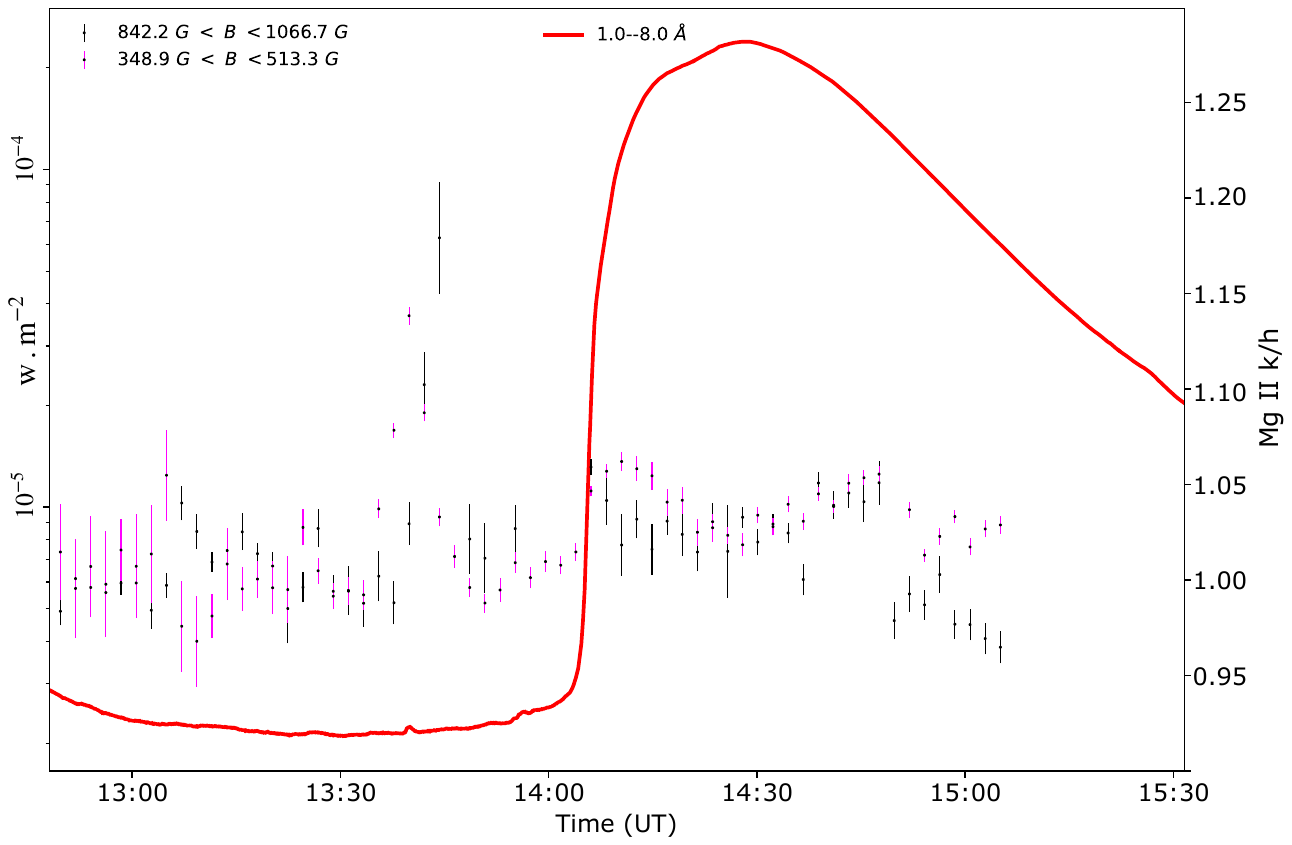}
    \caption{Time evolution of the \ion{Mg}{2} k to h line intensity ratio obtained from averaged spectrum over the corresponding magnetic flux bin as labeled for the Oct 22, 2014 flare. Over-plotted red solid line displays the 1{--}8~{\AA} GOES X-ray light curve.}
    \label{fig:op-dep-x}
\end{figure*}

Similar to the M-class class flare studied in the previous section, we derive the \ion{Mg}{2}~k to h line intensity ratio for various bins of the magnetic flux density within the flaring region and study their time evolution. In Fig.~\ref{fig:op-dep-x}, we plot the intensity ratio in black and magenta colors for two different magnetic field bins. In the same panel, we also over-plot, 1{--}8~{\AA} GOES light curve (red solid line). We note that, unlike the M-class flare, we do not find any detectable change in the intensity ratio during the flare. There are about two to three data points showing an enhancement in the ratio, but that is much before the start of the flare.

\subsection{C3.5 Flare Observed on Feb 03, 2015}
\begin{figure*}[ht!]
    \centering
\hspace*{-.6in}
\includegraphics[width=1.12\textwidth,trim={7cm 5cm 7.7cm 6cm},clip]{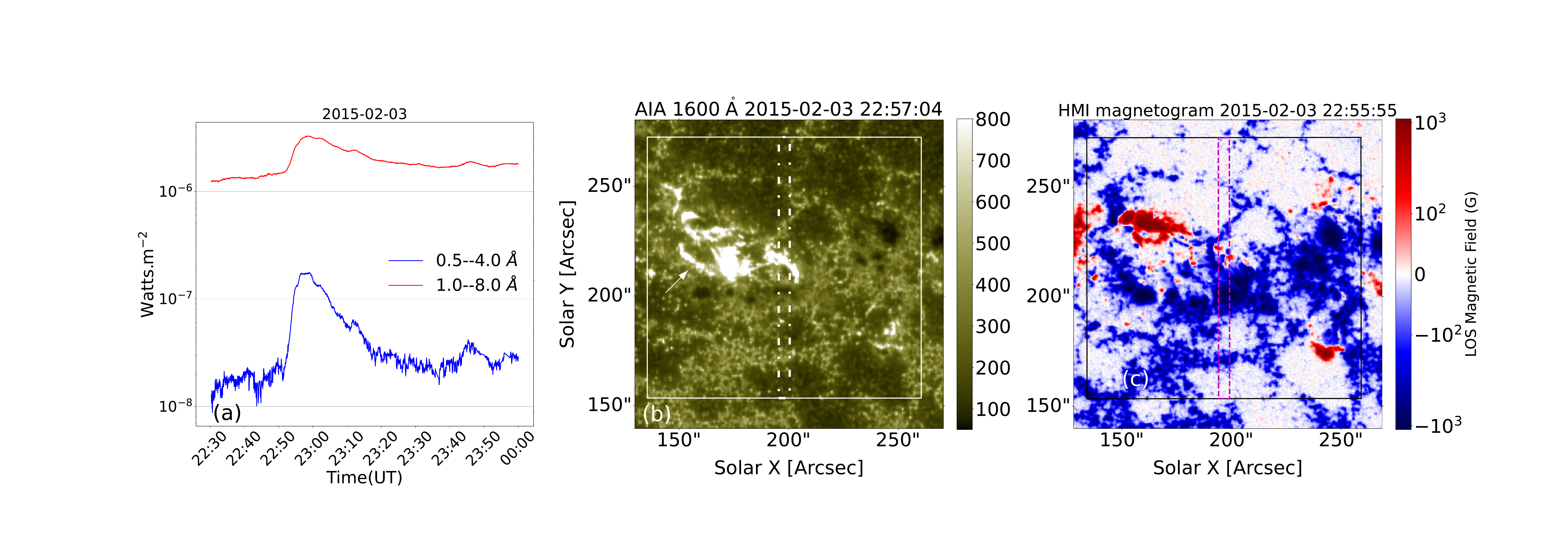}
\caption{The C class flare observed on February 3rd, 2015. Panel a: GOES flux plot in 0.5{--}4~{\AA} (blue) and 1.0{--}8.0~{\AA} (red). Panel b: AIA 1600~{\AA} image of the flaring region. Panel c: LOS magnetic flux density map obtained from HMI near the peak of the flare. The over-plotted white (black) boxes in panel b (c) represents the IRIS SJI FOV. The over-plotted white dot-dashed (magenta dashed) box in panel b (c) show the IRIS raster FOV.}\label{flare3}
\end{figure*}

On Feb 03, 2015, AR 12277 produced a C-class flare that peaked at 22:55~UT. Fig.~\ref{flare3} shows the GOES flux plot of the flare in 0.5{--}4~{\AA} (blue) and 1.0{--}8.0~{\AA} (red) in panel (a). Fig.~\ref{flare3}b \& c displays AIA 1600~{\AA} image and LOS magnetic flux density map obtained from HMI, respectively, recorded near the peak of the flare. 

\begin{figure}[ht!]
    \centering
\includegraphics[width=0.5\textwidth,trim={4cm 5cm 4cm 4cm},clip]{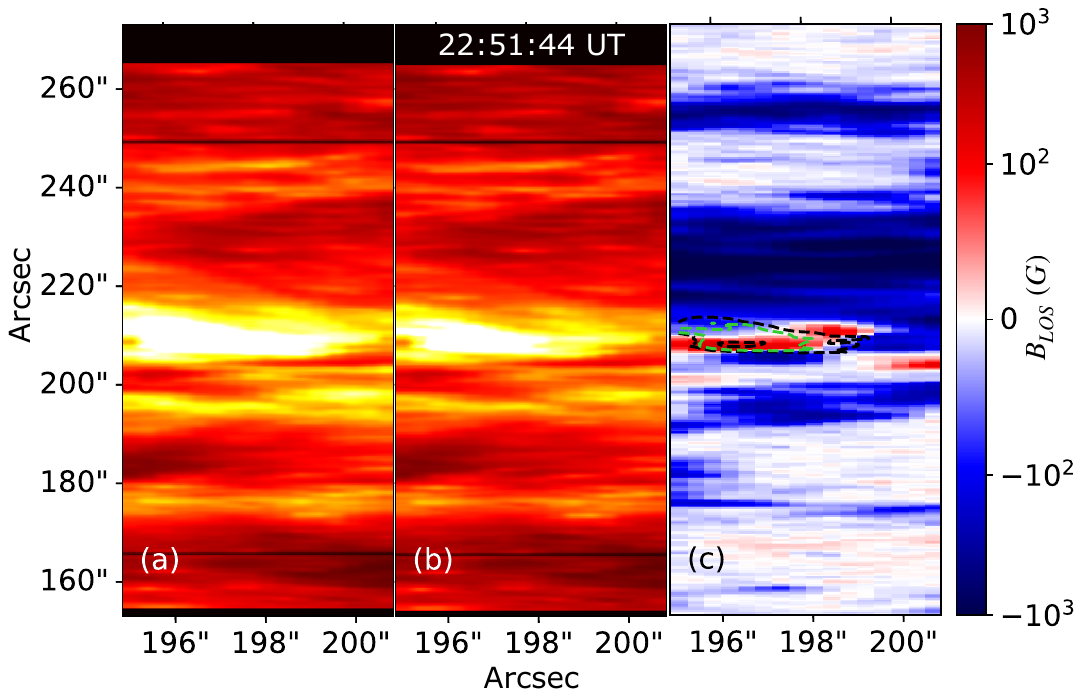}
\caption{Obtained intensity maps in \ion{Mg}{2}~k (panel a) and h (panel b) lines at the peak of the flare. The rastrered HMI LOS magnetic field density map is shown in panel c. The black and lime green contours on panel c show the contours of \ion{Mg}{2}~k (panel a) and h (panel b) intensity.}
\label{fig:align_raster_flare3}
\end{figure}

The over-plotted white (black) box in panel b (c) show the IRIS SJI FOV. The over-plotted white dot-dashed (magenta dashed) box in panel b (c) show the IRIS raster FOV. The IRIS observed this flare with a large dense 16 step raster with a FOV of 5"$\times$119" and step size of 0.35\arcsec. The flare exhibits a double ribbon structure. The east ribbon elongates earlier and merges into the west ribbon. The IRIS raster scans through the west edge of the west ribbon through its eruption and merging with the east ribbon. Fig.~\ref{fig:align_raster_flare3}, displays the intensity maps obtained in \ion{Mg}{2}~h \& k (panels a \& b). The rastered LOS magnetic field maps is shown in panel c. The black \& green dashed contours in the LOS magnetic field map (panel c) shows the intensity contours of \ion{Mg}{2}~h \& k (panels a \& b). 

\begin{figure}[ht!]
    \centering
    \includegraphics[trim={2cm 3cm 2cm 3cm},clip,width=0.5\textwidth]{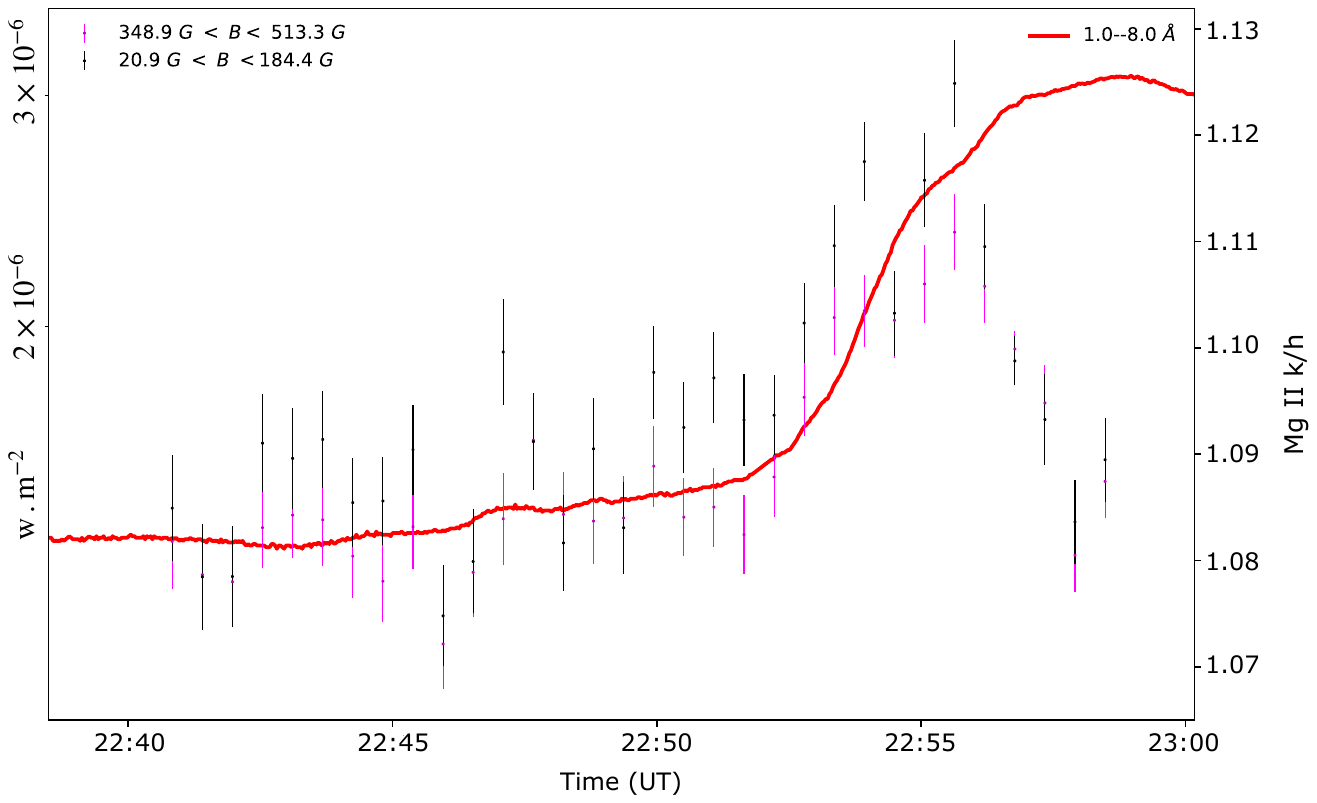}
    \caption{Time evolution of the \ion{Mg}{2} k to h line intensity ratio obtained from averaged spectrum over the corresponding magnetic flux bin as labeled for the Feb 3, 2015 flare. Over-plotted red solid line displays the 1{--}8~{\AA} GOES X-ray light curve.}
    \label{fig:optical_dep_ev_c}
\end{figure}

We plot the GOES 1{--}8~{\AA}~light curve (red solid line) along with the time evolution of the k to h line intensity ratio averaged within the bins of two different magnetic flux densities as a function of time in the Fig.~\ref{fig:optical_dep_ev_c}. The \ion{Mg}{2} k to h line intensity ratio shows a similar variation like the M class flare. It peaks during the impulsive phase (k/h$\sim$ 1.12), and reduces to preflare values (k/h$\sim$ 1.09) during the decay phase of the flare when the GOES light curve peaks. We note that the change in \ion{Mg}{2} k to h line intensity ratio is less prominent for this flare compared to the M-class event. However, we also note that the ratio shows a persistent increase from the start of the flare.

\section{Summary and Conclusions}

The \ion{Mg}{2}~k to h line intensity ratios is a useful diagnostic for the opacity of the local plasma. It may also help us to understand changes in the local plasma environment in the solar chromosphere during flares. In this paper, we have studied the time variation of \ion{Mg}{2}~k to h line intensity ratios during the evolution of three flares belonging to X-class, M-class and C-class. We have also studied the variation of intensity ratios in different magnetic flux density bins. For this purpose, we have used observations recorded by IRIS and HMI. For the co-alignment of IRIS and HMI observations, we have used 1600~{\AA} images recorded by AIA.

It is well known that the \ion{Mg}{2} profiles vary significantly in shape, spatially within the flaring region \citep{dalda23,panos18}. As we bin the IRIS observations depending on their magnetic field strength, it is worth mentioning that any variation in space is averaged out and not addressed in the analysis.

Our observations show that the \ion{Mg}{2}~k to h line intensity ratios change during flares. For the M-class and C-class flares, the ratio starts to increase at the time of the start of the flare, peaks approximately halfway in the impulsive phase and show a steep decline thereafter (see Figs.~\ref{fig:optical_dep_ev_m}~\&~\ref{fig:optical_dep_ev_c}). The ratios fall even below the pre-flare conditions during the later stage of the flare (peak and decline phase). However, we note that this behaviour is only observed in the M and C-class flares and not for the X-class flare.

Our observations did not show any correlation between the line intensity ratios and magnetic flux density. We show the line intensity ratio for different magnetic field strength in the Figs.~\ref{fig:optical_depth_m},~\ref{fig:optical_dep_ev_m},~\ref{fig:optical_dep_ev_c} illustrating the general behavior from weak to strong magnetic field strengths. This observation suggests that the magnetic field affects both the \ion{Mg}{2}~k and h lines similarly. Such effects cancel out while taking the ratios.

\cite{kerr15} studied \ion{Mg}{2}~k to h line intensity ratios and their time variation for individual pixels in quiet Sun region as well as flare locations. While they did not find any change in the ratios for the quiet Sun region, they noted that flaring pixels did show variations in the intensity ratios. It was speculated that this might be due to differences in heating conditions in the non flaring and flaring atmosphere. However, they did not observe any co-related change in the ratio with respect to the flare light curve. 

Our results are in line with those obtained by \cite{kerr15}, where we also find that the change in the ratio is highest during the impulsive phase of the flare and starts to decrease before the flare reaches its maximum in GOES 1{--}8~{\AA} observations. Such changes in the ratio may indicate variation in the optical depth of the local medium. Under this scenario, the optical depth first decreases during the impulsive heating phase and start to increase during the decay phase. 
While the decrease in the optical depth during the impulsive phase may be attributed to localized heating and chromospheric evaporation, the increase during the decay phase of the flare may be explained by condensation and down flows. We emphasize, however, that this interpretation is only a speculation, primarily because we do not observe such effects in the X-class flare.

We note that while the results obtained here for the M-class and C-class flare can be explained by the above discussed scenario, the result for the X-class is  more ambiguous. We could not develop a clear distinction in the general properties of these three flares. We note that while the C-class flare is confined, both M and X flares are eruptive flare. Therefore, we may suggest that in the X-class flare, the energy deposition is more impulsive and on a much shorter time scale than that could be sampled. The high impulsiveness may lead to a greater degree of ionization of the medium in comparison to the other two flares. This is also supported by the fact that there were a large number of saturated pixels in the data that were discarded. A firmer conclusion, however, requires analysis of more such flares, including numerical and theoretical modeling.

\begin{acknowledgments}
We thank the anonymous reviewer for the careful reading of our manuscript and providing us with many insightful comments and suggestions. We thank Peter Young for the valuable comments on the manuscript. IRIS is a NASA small explorer mission developed and operated by LMSAL with mission operations executed at NASA Ames Research Center and major contributions to downlink communications funded by ESA and the Norwegian Space Centre. The AIA data used here are courtesy of SDO (NASA) and the AIA consortium. SR is supported by the NASA HSO-Connect program, grant number 80NSSC20K1283. This research used version 4.1.5 \citep{sunpy_ver} of the SunPy open-source software package \citep{sunpy20} and PYTHON packages NumPy \citep{numpy}, Matplotlib \citep{matpltolib}, SciPy \citep{scipy}. 
\end{acknowledgments}

\appendix

\section{Comparison of methods used to obtain \ion{Mg}{2} intensities}\label{appendix}

\begin{figure}
    \centering
    \includegraphics[trim={1cm 0.5cm 0.5cm 2cm},clip,width=0.8\textwidth]{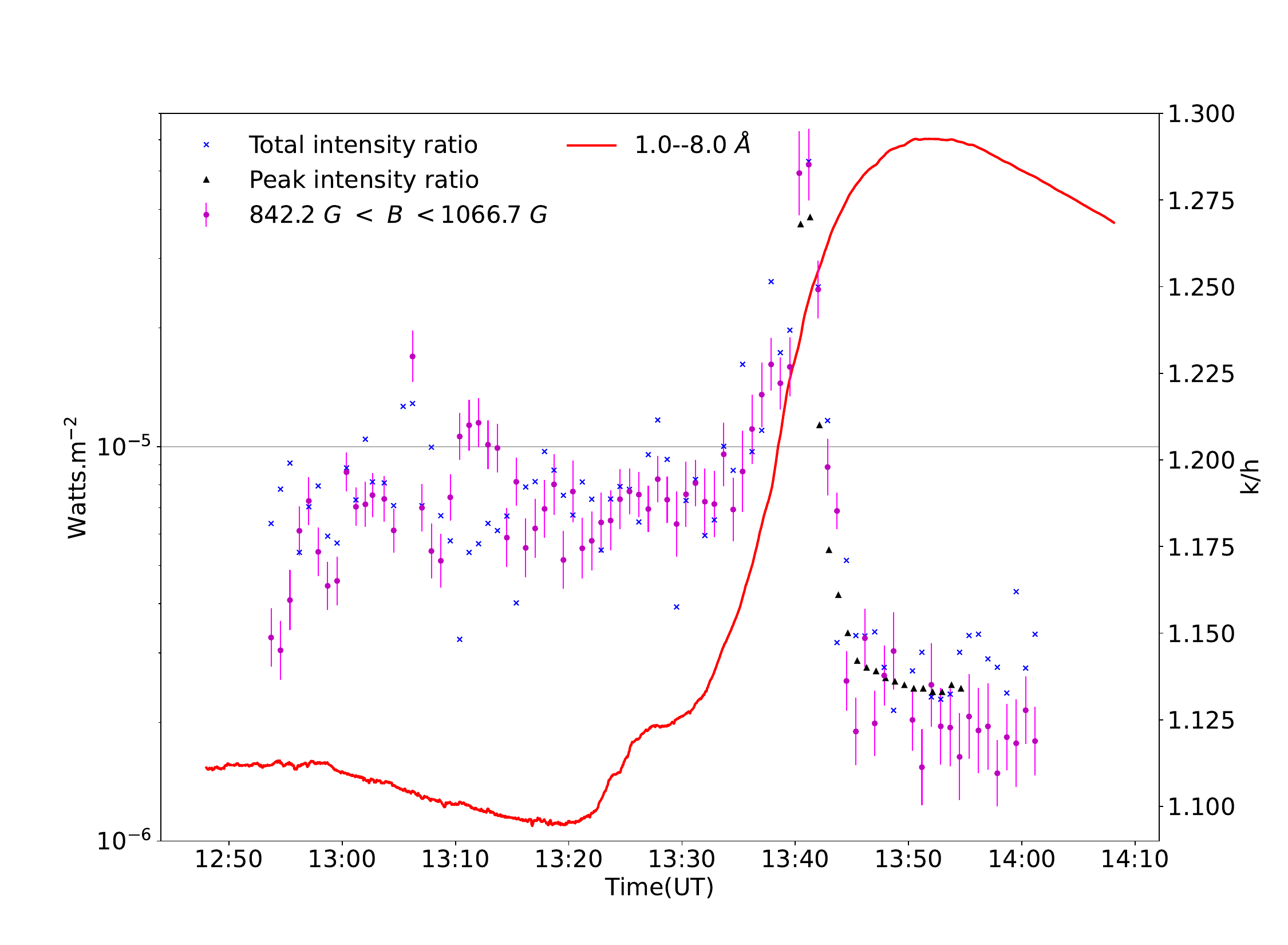}
    \caption{1{--}8~{\AA} GOES light curve of the flare over-plotted with the time variation of \ion{Mg}{2} k/h line intensity of the line intensity ratios for the binned spectra for the pixels in 842.2 G $<B<$ 1066.7 G(magenta points). Similarly, the ratio of the line intensities without fitting the line profiles is over plotted (blue cross). The black triangles are the peak intensity ratio where the line is single peaked.}. 
    \label{fig:mthd_comp}
    \end{figure}

There are a few methods to characterize the properties of the IRIS \ion{Mg}{2} lines. One of the key ones is to calculate the normalized cumulative distribution functions(CDFs) of the Mg lines, and use the various quartiles to quantify the line parameters like line centroid, line width and asymmetry of the line profiles. The \ion{Mg}{2} lines are integrated within a fixed wavelength range to calculate the CDFs. We compare our fitting method, with integrating the IRIS line intensity in a fixed wavelength window 279.4{--}279.8~{\AA} (\ion{Mg}{2} k) and 280.15{--}280.55~\AA(\ion{Mg}{2} h) in fig~\ref{fig:mthd_comp}, and with peak intensity ratios where the lines are single peaked. The goal of this analysis was to quantify if our fitting method shows any difference from integrating through the line profiles, or taking the ratio of peak intensity from single peaked profiles.

The red solid line is the GOES soft X-ray intensity in 1{--}8~{\AA}. The \ion{Mg}{2} k/h line intensity ratio for the binned spectra of magnetic field strength $842.2~G<B<1066.7~G$ is plotted with magenta circles. The same is plotted in  Fig.~\ref{fig:optical_depth_m}. The \ion{Mg}{2} k/h line intensity ratio for the same binned spectra obtained from integrating the IRIS observation in a fixed wavelength window, is plotted with blue crosses. The \ion{Mg}{2} k/h peak intensity ratio, when the k and h lines are single peaked, is shown with the black triangles. Both cases, the trend of the line intensity ratio behaves similarly as described earlier. It increases dramatically during the impulsive phase of the flare, and after the soft X-ray peak it decreases again to preflare values. The peak intensity ratio of the single peaked line profiles behave more monotonic during the decay phase of the flare.

\bibliography{mybib}{}

\begin{thebibliography}{}
\expandafter\ifx\csname natexlab\endcsname\relax\def\natexlab#1{#1}\fi
\providecommand{\url}[1]{\href{#1}{#1}}
\providecommand{\dodoi}[1]{doi:~\href{http://doi.org/#1}{\nolinkurl{#1}}}
\providecommand{\doeprint}[1]{\href{http://ascl.net/#1}{\nolinkurl{http://ascl.net/#1}}}
\providecommand{\doarXiv}[1]{\href{https://arxiv.org/abs/#1}{\nolinkurl{https://arxiv.org/abs/#1}}}

\bibitem[{{De Pontieu} {et~al.}(2014){De Pontieu}, {Title}, {Lemen}, {Kushner},
  {Akin}, {Allard}, {Berger}, {Boerner}, {Cheung}, {Chou}, {Drake}, {Duncan},
  {Freeland}, {Heyman}, {Hoffman}, {Hurlburt}, {Lindgren}, {Mathur}, {Rehse},
  {Sabolish}, {Seguin}, {Schrijver}, {Tarbell}, {W{\"u}lser}, {Wolfson},
  {Yanari}, {Mudge}, {Nguyen-Phuc}, {Timmons}, {van Bezooijen}, {Weingrod},
  {Brookner}, {Butcher}, {Dougherty}, {Eder}, {Knagenhjelm}, {Larsen},
  {Mansir}, {Phan}, {Boyle}, {Cheimets}, {DeLuca}, {Golub}, {Gates}, {Hertz},
  {McKillop}, {Park}, {Perry}, {Podgorski}, {Reeves}, {Saar}, {Testa}, {Tian},
  {Weber}, {Dunn}, {Eccles}, {Jaeggli}, {Kankelborg}, {Mashburn}, {Pust},
  {Springer}, {Carvalho}, {Kleint}, {Marmie}, {Mazmanian}, {Pereira}, {Sawyer},
  {Strong}, {Worden}, {Carlsson}, {Hansteen}, {Leenaarts}, {Wiesmann},
  {Aloise}, {Chu}, {Bush}, {Scherrer}, {Brekke}, {Martinez-Sykora}, {Lites},
  {McIntosh}, {Uitenbroek}, {Okamoto}, {Gummin}, {Auker}, {Jerram}, {Pool}, \&
  {Waltham}}]{IRIS}
{De Pontieu}, B., {Title}, A.~M., {Lemen}, J.~R., {et~al.} 2014, \solphys, 289,
  2733, \dodoi{10.1007/s11207-014-0485-y}

\bibitem[{{Fletcher}(2010)}]{fletcher10}
{Fletcher}, L. 2010, \memsai, 81, 616.
\newblock \doarXiv{1001.0739}

\bibitem[{Harris {et~al.}(2020)Harris, Millman, van~der Walt, Gommers,
  Virtanen, Cournapeau, Wieser, Taylor, Berg, Smith, Kern, Picus, Hoyer, van
  Kerkwijk, Brett, Haldane, del R{\'{i}}o, Wiebe, Peterson,
  G{\'{e}}rard-Marchant, Sheppard, Reddy, Weckesser, Abbasi, Gohlke, \&
  Oliphant}]{numpy}
Harris, C.~R., Millman, K.~J., van~der Walt, S.~J., {et~al.} 2020, Nature, 585,
  357, \dodoi{10.1038/s41586-020-2649-2}

\bibitem[{Hunter(2007)}]{matpltolib}
Hunter, J.~D. 2007, Computing in Science \& Engineering, 9, 90,
  \dodoi{10.1109/MCSE.2007.55}

\bibitem[{{Karlick{\'y}} {et~al.}(2018){Karlick{\'y}}, {Zemanov{\'a}},
  {Dud{\'\i}k}, \& {Radziszewski}}]{karlick18}
{Karlick{\'y}}, M., {Zemanov{\'a}}, A., {Dud{\'\i}k}, J., \& {Radziszewski}, K.
  2018, \apjl, 854, L29, \dodoi{10.3847/2041-8213/aaadf9}

\bibitem[{{Kerr} {et~al.}(2015){Kerr}, {Sim{\~o}es}, {Qiu}, \&
  {Fletcher}}]{kerr15}
{Kerr}, G.~S., {Sim{\~o}es}, P.~J.~A., {Qiu}, J., \& {Fletcher}, L. 2015, \aap,
  582, A50, \dodoi{10.1051/0004-6361/201526128}

\bibitem[{{Leenaarts} {et~al.}(2013{\natexlab{a}}){Leenaarts}, {Pereira},
  {Carlsson}, {Uitenbroek}, \& {De Pontieu}}]{leenarts13a}
{Leenaarts}, J., {Pereira}, T.~M.~D., {Carlsson}, M., {Uitenbroek}, H., \& {De
  Pontieu}, B. 2013{\natexlab{a}}, \apj, 772, 90,
  \dodoi{10.1088/0004-637X/772/2/90}

\bibitem[{{Leenaarts} {et~al.}(2013{\natexlab{b}}){Leenaarts}, {Pereira},
  {Carlsson}, {Uitenbroek}, \& {De Pontieu}}]{leenarts13b}
---. 2013{\natexlab{b}}, \apj, 772, 89, \dodoi{10.1088/0004-637X/772/2/89}

\bibitem[{{Lemaire} {et~al.}(1984){Lemaire}, {Choucq-Bruston}, \&
  {Vial}}]{lemaire84}
{Lemaire}, P., {Choucq-Bruston}, M., \& {Vial}, J.~C. 1984, \solphys, 90, 63,
  \dodoi{10.1007/BF00153785}

\bibitem[{{Lemen} {et~al.}(2012){Lemen}, {Title}, {Akin}, {Boerner}, {Chou},
  {Drake}, {Duncan}, {Edwards}, {Friedlaender}, {Heyman}, {Hurlburt}, {Katz},
  {Kushner}, {Levay}, {Lindgren}, {Mathur}, {McFeaters}, {Mitchell}, {Rehse},
  {Schrijver}, {Springer}, {Stern}, {Tarbell}, {Wuelser}, {Wolfson}, {Yanari},
  {Bookbinder}, {Cheimets}, {Caldwell}, {Deluca}, {Gates}, {Golub}, {Park},
  {Podgorski}, {Bush}, {Scherrer}, {Gummin}, {Smith}, {Auker}, {Jerram},
  {Pool}, {Soufli}, {Windt}, {Beardsley}, {Clapp}, {Lang}, \& {Waltham}}]{aia}
{Lemen}, J.~R., {Title}, A.~M., {Akin}, D.~J., {et~al.} 2012, \solphys, 275,
  17, \dodoi{10.1007/s11207-011-9776-8}

\bibitem[{{Levens} \& {Labrosse}(2019)}]{levens19}
{Levens}, P.~J., \& {Labrosse}, N. 2019, \aap, 625, A30,
  \dodoi{10.1051/0004-6361/201833132}

\bibitem[{{Li} {et~al.}(2017){Li}, {Zhang}, \& {Hou}}]{li17}
{Li}, T., {Zhang}, J., \& {Hou}, Y. 2017, \apj, 848, 32,
  \dodoi{10.3847/1538-4357/aa8c01}

\bibitem[{Mariska(1992)}]{mariska92}
Mariska, J.~T. 1992, The solar transition region, Cambridge astrophysics series
  ; 23 (Cambridge [England] ; New York: Cambridge University Press)

\bibitem[{{Milligan} {et~al.}(2014){Milligan}, {Kerr}, {Dennis}, {Hudson},
  {Fletcher}, {Allred}, {Chamberlin}, {Ireland}, {Mathioudakis}, \&
  {Keenan}}]{milligan14}
{Milligan}, R.~O., {Kerr}, G.~S., {Dennis}, B.~R., {et~al.} 2014, \apj, 793,
  70, \dodoi{10.1088/0004-637X/793/2/70}

\bibitem[{Mumford {et~al.}(2023)Mumford, Freij, Stansby, Christe, Ireland,
  Mayer, Shih, Hughitt, Ryan, Liedtke, Hayes, Pérez-Suárez, I., Barnes,
  Chakraborty, Inglis, Pattnaik, Sipőcz, MacBride, Sharma, Leonard, Hewett,
  Hamilton, Manhas, Panda, Earnshaw, Choudhary, Kumar, Singh, Chanda, Haque,
  Kirk, Mueller, Konge, Srivastava, Wentzel-Long, Jain, Bennett, Baruah,
  Arbolante, Charlton, Maloney, Mishra, Paul, Verma, Chorley, Chouhan,
  Himanshu, Mason, Zivadinovic, Modi, Sharma, Naman9639, Bobra, Rozo, Manley,
  Ivashkiv, Laitinen, Chatterjee, von Forstner, Bazán, Stern, Gieseler, Evans,
  Jain, Malocha, Ghosh, Airmansmith97, Stańczak, Singh, Visscher, Verma,
  SophieLemos, Agrawal, Alam, Buddhika, Pathak, Rideout, Sharma, Park, Bates,
  Wilson, Shukla, Giger, Mishra, Sharma, Goel, Taylor, Cetusic, Reiter, Jacob,
  Inchaurrandieta, Dacie, Dubey, Eigenbrot, Bray, Surve, Zahniy, Sidhu,
  Meszaros, Parkhi, Russell, Bose, Pandey, Price-Whelan, J, Chicrala, Ankit,
  Guennou, D'Avella, Williams, Verma, Ballew, Agrawal, Murphy, Lodha,
  Robitaille, Augspurger, Krishan, honey, neerajkulk, Bhope, Gaba, Hill,
  Mampaey, Wiedemann, Molina, Briseno, Keşkek, Habib, Letts, Singaravelan,
  Ranjan, Altunian, Streicher, Gomillion, Agarwal, Kothari, Nomiya,
  mridulpandey, Stevens, B, Bahuleyan, Kaszynski, W, Mehrotra, Tang, Sinha,
  Smith, Kustov, Stone, Bard, Arias, Tollerud, Dover, Verstringe, Kumar,
  Mathur, Babuschkin, Calixto, Wimbish, Qing, Buitrago-Casas, Krishna,
  Chaudhari, Hiware, Ghosh, Lyes, Mangaonkar, Cheung, Mendero, Dedhia,
  Schoentgen, Shahdadpuri, Srinivasan, Gyenge, Mekala, Das, Mishra, Sharma,
  Srikanth, Jain, Kannojia, Yadav, Paul, Wilkinson, Caswell, Braccia, Pereira,
  Gates, Dang, Bankar, Jamieson, Agrawal, platipo, resakra, tal66, yasintoda,
  Attie, \& Murray}]{sunpy_ver}
Mumford, S.~J., Freij, N., Stansby, D., {et~al.} 2023, SunPy, v4.1.5,  Zenodo,
  \dodoi{10.5281/zenodo.7850372}

\bibitem[{{Panos} \& {Kleint}(2021{\natexlab{a}})}]{panos21}
{Panos}, B., \& {Kleint}, L. 2021{\natexlab{a}}, \apj, 915, 77,
  \dodoi{10.3847/1538-4357/ac00c0}

\bibitem[{{Panos} \& {Kleint}(2021{\natexlab{b}})}]{panos21_2}
---. 2021{\natexlab{b}}, \apj, 915, 77, \dodoi{10.3847/1538-4357/ac00c0}

\bibitem[{{Panos} {et~al.}(2018){Panos}, {Kleint}, {Huwyler}, {Krucker},
  {Melchior}, {Ullmann}, \& {Voloshynovskiy}}]{panos18}
{Panos}, B., {Kleint}, L., {Huwyler}, C., {et~al.} 2018, \apj, 861, 62,
  \dodoi{10.3847/1538-4357/aac779}

\bibitem[{{Patsourakos} {et~al.}(2004){Patsourakos}, {Antiochos}, \&
  {Klimchuk}}]{patsourakos04}
{Patsourakos}, S., {Antiochos}, S.~K., \& {Klimchuk}, J.~A. 2004, \apj, 614,
  1022, \dodoi{10.1086/423779}

\bibitem[{{Pereira} {et~al.}(2015){Pereira}, {Carlsson}, {De Pontieu}, \&
  {Hansteen}}]{pereira15}
{Pereira}, T. M.~D., {Carlsson}, M., {De Pontieu}, B., \& {Hansteen}, V. 2015,
  \apj, 806, 14, \dodoi{10.1088/0004-637X/806/1/14}

\bibitem[{{Pereira} {et~al.}(2013){Pereira}, {Leenaarts}, {De Pontieu},
  {Carlsson}, \& {Uitenbroek}}]{pereira13}
{Pereira}, T.~M.~D., {Leenaarts}, J., {De Pontieu}, B., {Carlsson}, M., \&
  {Uitenbroek}, H. 2013, \apj, 778, 143, \dodoi{10.1088/0004-637X/778/2/143}

\bibitem[{{Pesnell} {et~al.}(2012){Pesnell}, {Thompson}, \& {Chamberlin}}]{sdo}
{Pesnell}, W.~D., {Thompson}, B.~J., \& {Chamberlin}, P.~C. 2012, \solphys,
  275, 3, \dodoi{10.1007/s11207-011-9841-3}

\bibitem[{{Polito} {et~al.}(2023){Polito}, {Kerr}, {Xu}, {Sadykov}, \&
  {Lorincik}}]{polito23}
{Polito}, V., {Kerr}, G.~S., {Xu}, Y., {Sadykov}, V.~M., \& {Lorincik}, J.
  2023, \apj, 944, 104, \dodoi{10.3847/1538-4357/acaf7c}

\bibitem[{{Sainz Dalda} \& {De Pontieu}(2023)}]{dalda23}
{Sainz Dalda}, A., \& {De Pontieu}, B. 2023, Frontiers in Astronomy and Space
  Sciences, 10, 1133429, \dodoi{10.3389/fspas.2023.1133429}

\bibitem[{{Scherrer} {et~al.}(2012){Scherrer}, {Schou}, {Bush}, {Kosovichev},
  {Bogart}, {Hoeksema}, {Liu}, {Duvall}, {Zhao}, {Title}, {Schrijver},
  {Tarbell}, \& {Tomczyk}}]{hmi}
{Scherrer}, P.~H., {Schou}, J., {Bush}, R.~I., {et~al.} 2012, \solphys, 275,
  207, \dodoi{10.1007/s11207-011-9834-2}

\bibitem[{{Sharma} {et~al.}(2016){Sharma}, {Tripathi}, {Isobe}, \&
  {Ghosh}}]{sharma16}
{Sharma}, R., {Tripathi}, D., {Isobe}, H., \& {Ghosh}, A. 2016, \apj, 823, 47,
  \dodoi{10.3847/0004-637X/823/1/47}

\bibitem[{{Spirock} {et~al.}(2002){Spirock}, {Yurchyshyn}, \&
  {Wang}}]{spirock02}
{Spirock}, T.~J., {Yurchyshyn}, V.~B., \& {Wang}, H. 2002, \apj, 572, 1072,
  \dodoi{10.1086/340431}

\bibitem[{{The SunPy Community} {et~al.}(2020){The SunPy Community}, Barnes,
  Bobra, Christe, Freij, Hayes, Ireland, Mumford, Perez-Suarez, Ryan, Shih,
  Chanda, Glogowski, Hewett, Hughitt, Hill, Hiware, Inglis, Kirk, Konge, Mason,
  Maloney, Murray, Panda, Park, Pereira, Reardon, Savage, Sipőcz, Stansby,
  Jain, Taylor, Yadav, Rajul, \& Dang}]{sunpy20}
{The SunPy Community}, Barnes, W.~T., Bobra, M.~G., {et~al.} 2020, The
  Astrophysical Journal, 890, 68, \dodoi{10.3847/1538-4357/ab4f7a}

\bibitem[{{Tripathi} {et~al.}(2004){Tripathi}, {Bothmer}, \&
  {Cremades}}]{TriBC_2004}
{Tripathi}, D., {Bothmer}, V., \& {Cremades}, H. 2004, \aap, 422, 337,
  \dodoi{10.1051/0004-6361:20035815}

\bibitem[{{van Regemorter}(1962)}]{henri62}
{van Regemorter}, H. 1962, \apj, 136, 906, \dodoi{10.1086/147445}

\bibitem[{Virtanen {et~al.}(2020)Virtanen, Gommers, Oliphant, Haberland, Reddy,
  Cournapeau, Burovski, Peterson, Weckesser, Bright, {van der Walt}, Brett,
  Wilson, Millman, Mayorov, Nelson, Jones, Kern, Larson, Carey, Polat, Feng,
  Moore, {VanderPlas}, Laxalde, Perktold, Cimrman, Henriksen, Quintero, Harris,
  Archibald, Ribeiro, Pedregosa, {van Mulbregt}, \& {SciPy 1.0
  Contributors}}]{scipy}
Virtanen, P., Gommers, R., Oliphant, T.~E., {et~al.} 2020, Nature Methods, 17,
  261, \dodoi{10.1038/s41592-019-0686-2}

\bibitem[{{Wang} {et~al.}(2002){Wang}, {Spirock}, {Qiu}, {Ji}, {Yurchyshyn},
  {Moon}, {Denker}, \& {Goode}}]{wang02}
{Wang}, H., {Spirock}, T.~J., {Qiu}, J., {et~al.} 2002, \apj, 576, 497,
  \dodoi{10.1086/341735}

\bibitem[{{Ye} {et~al.}(2016){Ye}, {Liu}, \& {Wang}}]{dandan16}
{Ye}, D.-D., {Liu}, C., \& {Wang}, H. 2016, Research in Astronomy and
  Astrophysics, 16, 95, \dodoi{10.1088/1674-4527/16/6/095}

\end{thebibliography}
\bibliographystyle{aasjournal}

\end{document}